\documentclass[apl,superscriptaddress,reprint,twocolumn,showpacs]{revtex4-1}

\usepackage{graphicx}
\usepackage{dcolumn}
\usepackage{bm}
\usepackage[mathlines]{lineno}
\usepackage{natbib}
\usepackage{color}

\begin{document}

\preprint{AIP/123-QED}


\title[Trapping and manipulation of individual nanoparticles in a planar Paul trap]{Trapping and manipulation of individual nanoparticles in a planar Paul trap}

\author{I. Alda}
\altaffiliation{These authors contributed equally}
\affiliation{ICFO-Institut de Ciencies Fotoniques, The Barcelona Institute of Science and Technology, 08860 Castelldefels (Barcelona), Spain}

\author{J. Berthelot $^{*}$}
 \email{johann.berthelot@fresnel.fr}
\altaffiliation[Current address: ]{CNRS, Aix-Marseille Université, Centrale Marseille, Institut Fresnel, UMR 7249, 13013 Marseille, France}
\affiliation{ICFO-Institut de Ciencies Fotoniques, The Barcelona Institute of Science and Technology, 08860 Castelldefels (Barcelona), Spain}

\author{R. A. Rica}%
\affiliation{ICFO-Institut de Ciencies Fotoniques, The Barcelona Institute of Science and Technology, 08860 Castelldefels (Barcelona), Spain}

\author{R. Quidant}
\email{romain.quidant@icfo.es}
\affiliation{ICFO-Institut de Ciencies Fotoniques, The Barcelona Institute of Science and Technology, 08860 Castelldefels (Barcelona), Spain}
\affiliation{ICREA --- Instituci\'o Catalana de Recerca i Estudis Avan\c{c}ats, 08010 Barcelona, Spain}

\begin{abstract}
\noindent
Visualisation and manipulation of nanoscale matter is one of the main and current challenges in nanosciences. To this aim, different techniques have been recently developed to non-invasively trap and manipulate nano-specimens, like nanoparticles or molecules. However, operating in air or vacuum still remains very challenging since most approaches are limited to a liquid environment. In this letter, we design and characterise a planar Paul trap optimised to trap and manipulate individual charged nanoparticles. This configuration offers competitive capabilities to manipulate nano-specimens in air or vacuum including in-plane integration, high trap confinement along with dynamical trap reconfiguration.  
\end{abstract}

                          
\maketitle

Current research in nanotechnology demands tools to accurately and non-invasively manipulate objects at the nanoscale. Conventional optical tweezers (OTs) have these capabilities for micrometer sized objects and neutral atoms \cite{Ashkin1986}, but are inadequate for the intermediate size range (i.e. 1-100 nm size objects), mainly due to the strong decrease of optical forces with the particle's volume\cite{Neuman}. As an alternative, researchers have developed near field optical trapping schemes based on nanoplasmonics \cite{Juan2011} and silicon photonics \cite{Erickson2011}. Such approaches allow stable trapping from  single proteins \cite{Pang2011a} up to nanoparticles of a few tens of nanometers \cite{Juan2009, Pang2011, Chen2011} and have recently allowed 3D manipulation \cite{Berthelot2014}. Other alternatives are surface electrostatic traps \cite{kim2014scanning}  and the Anti-Brownian Eletrophoretic trap (ABEL) \cite{cohen2005method, Kayci2015}. However, all remain limited to a liquid environment and it still represents an important challenge to adapt this technology to other mediums, i.e. air or vacuum. Interestingly, the optical levitation of nanoparticles in high vacuum has appeared as an exquisite platform for several applications in force sensing and optomechanics~\cite{li2011millikelvin,gieseler2012subkelvin,mestres2015cooling,ranjit2015attonewton,jain2016direct}. However, photothermal damage strongly limits the nanoparticle's material, and typically only silica (very low absorption) is used. Yet, it would be very interesting to levitate other nanoparticles featuring internal energy level like colour centers~\cite{neukirch2015multi,hoang2015observation,rahman2015burning}.

Originally used to trap ions, Paul traps (PTs)\cite{Paul1990} can levitate charged nano-objects \cite{Kuhlicke2014,howder2014optically,bell2014single,Millen2015, Kuhlicke2015, Nagornykh2015, Tanaka2014}. PTs found applications in various fields in physics \cite{Major2005, Werth2009},  chemistry \cite{Gerlich2007, Willitsch2008} and even biology \cite{Ostendorf2006, Offenberg2009}. They are based on a quadrupole potential originating from a RF time varying electrical field created by a set of properly arranged electrodes. The trajectory of the trapped object is described by the superposition of a fast driven oscillation at the frequency of the RF field called micromotion and a slower one called macromotion \cite{Berkeland1998}. PTs electrodes designs can be challenging to fabricate and are usually bulky, hence complicating the trap loading and its optical interrogation. An interesting alternative to standard PT are planar PTs (PPTs) whose lithographed electrodes are integrated in a single plane \cite{Chiaverini2005, Kumph2016}. Their compatibility with Printed Circuit Board (PCB) technology makes their fabrication simple and convenient \cite{Kim2010, MuirKumph2011, Tanaka2014, Maurice2015}.

\begin{figure}[hbt]
    \includegraphics[scale=0.31]{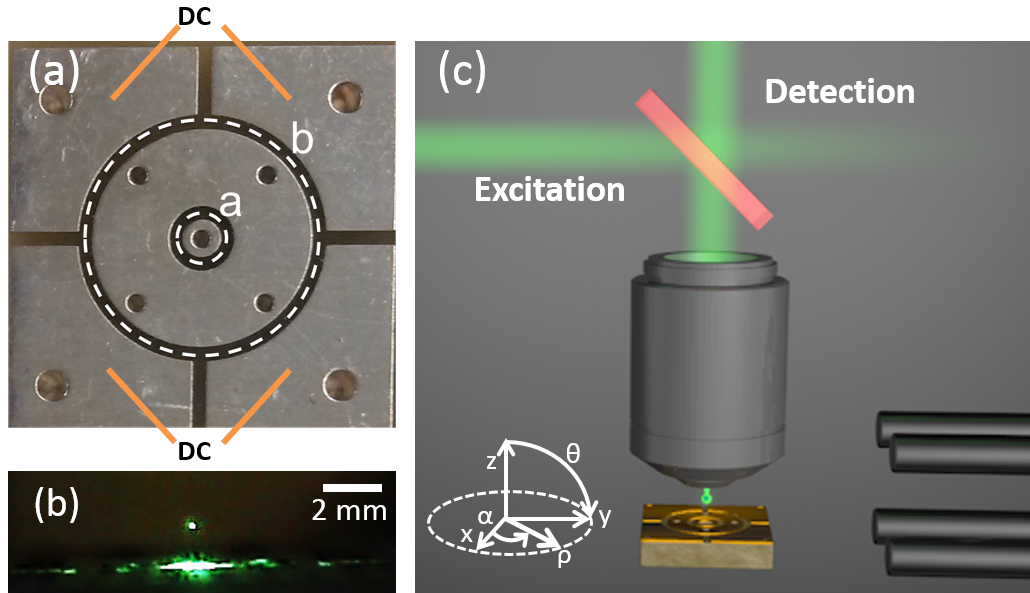}
\caption{(a) Image of our planar Paul trap (PPT) on PCB. The  radius of the inner and outer electrodes are $a$ and $b$, respectively. Our design includes compensation DC electrodes on the corners. All connections are on the backside of the PCB.  (b) Levitating 100 nm dielectric nanoparticle trapped with a PPT. Both images in (a) and (b) have the same scale. (c) Sketch of the setup to measure the PPT's confinement. A 532 nm laser is focused through a 50x objective (0.5 NA).  The scattering light is collected by the same objective for monitoring the nanoparticle motion. }
\label{fig:PCB}
\end{figure}

In this letter, we report an optimized Planar Paul Trap (PPT) for 3D manipulation of single charged polystyrene nanoparticles under ambient and atmospheric conditions. The trap enables stable confinement of the nanoparticle for days. Full manipulation capabilities are demonstrated by rotating the trap and modifying the trapping distance to the surface.  We also identify a design that maximizes spatial confinement and we use position quadrant photodetection (QPD) to characterize the 3D dynamics of the trapped nanoparticle.

Our PPT on PCB is formed by two concentric electrodes with radii $a$ (inner electrode) and $b$ (outer electrode), and four corner compensation DC electrodes as shown on Fig. \ref{fig:PCB}(a). This design features an empty central hole, providing an optical access to the trapped specimen \cite{VanDevender2010}. The electrical connections are made from the backside of the PCB. Both sides are connected via vertical conducting holes filled with alloy, keeping the topside clear for manipulation.  To start with, we use a similar design to the one presented in ref. \cite{Kim2010} with dimensions $a = 1.07$ mm and $b = 3.62$ mm (labeled as PPT$_1$ in the following). We used a suspension of 100 nm diameter polymer spheres dispersed in ethanol. The particles are charged with electrospray \cite{Gaskell1997} and guided to the PPT with a linear Paul trap at an amplitude of 1000 V and a frequency of 800 Hz  \cite{Kim2010,Gregor2009}. The PPT is powered by two wave function generators and two high voltage amplifiers. In the simplest operation mode, only the outer electrode is driven by the RF signal (amplitude of 340 V,  frequency of 4 kHz) while keeping the inner electrode grounded.  Figure \ref{fig:PCB}(b) shows a side view scattering picture from a single particle levitating in the PPT. A four axis stage ($x$, $y$, $z$ and $\theta$) moves the PCB and a 532 nm laser interrogates the trapped object. This light is focused to a 1.23 $\mu$m diameter spot size using a long working distance $50\times$ objective (NA= 0.5). The collected back-scattered light propagates through a 45:55 beam splitter and is focused onto a CCD camera (see Fig.  \ref{fig:PCB} (c)).

\begin{figure}[hbt]
    \includegraphics[scale=0.55]{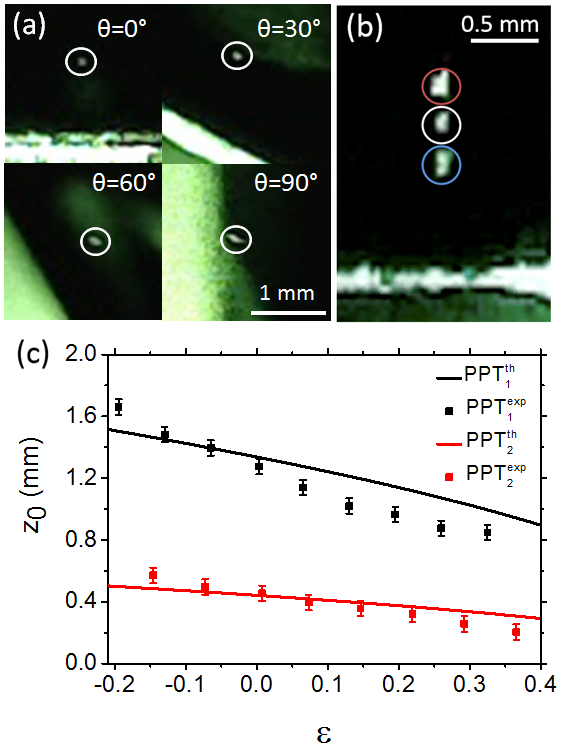}
\caption{(a) Mosaic of images taken from the side showing the manipulation of a single trapped nanoparticle with the PPT (0$^{\circ}$ to 90$^{\circ}$). We can also rotate it 180$^\circ$. (b) Superimposed images for three different $\epsilon$ values: $\epsilon = 0$ in white, $\epsilon>0$ in blue, and $\epsilon<0$ in red (see text). (c) Experimental data for $z_0$ for different values of $\epsilon$ for both PCBs (points) and the corresponding theoretical predictions (solid lines).}
\label{fig:height}
\end{figure}

The PPT compactness along with its high trapping stability enables unique manipulation capabilities that could not be achieved with optical tweezers. For instance, we demonstrate in Fig. \ref{fig:height}(a) that the nanoparticle is kept trapped while rotating the PPT around its $x$ axis. Further control over the trap features can be reached by applying an additional RF signal $V_{in}$ of the same frequency (4~kHz) to the inner circular electrode. By varying $\epsilon = V_{in}/V_{out}$ \cite{Kim2010} we are able to adjust the trapping height (see Eq.~\ref{eq:z0}) \cite{Wesenberg2008, Kim2010}. Depending on the sign of $\epsilon$, which relates to the phase difference between the two driving fields, the trapping height decreases ($\epsilon>0$) or increases ($\epsilon<0$) as shown on Fig. \ref{fig:height} (b). The experimental dependence with $\epsilon$ of the particle height, plotted in graph Fig. \ref{fig:height} (c) for two PPT designs, shows that it can be changed by up to 100 $\%$.

The studied PPT geometry can be theoretically described using the model of reference \cite{Kim2010}. We consider the general case where the inner and outer electrode are powered with two RF fields of the same frequency. Using this model, we have optimised the trapping potential parameters in order to achieve both high confinement while maintaining PCB fabrication compatibility. Given the cylindrical symmetry of our design, the electric potential created by our PPT can be properly described in cylindrical coordinates $\alpha= \tan^{-1}{\left(y/x\right)}$,  $\rho = \sqrt{x^2+y^2}$, and $z$. Trapping occurs at $\rho = 0$, resulting in the potential depending only on $z$. For a single trapped particle, its motion is typically described by a pseudopotential $\Psi$ that accounts for the macromotion. In the absence of defects or a DC bias it is expressed as \cite{Kim2010}:

\begin{equation}
\Psi(z,\rho,\epsilon) = A\mid\nabla\kappa(z,\rho,\epsilon) \mid^2, \\
\label{eq:pseudopot}
\end{equation}
where $A= \frac{Q^2V_{out}^2}{4M\Omega^2}$, $M$ and $Q$ are the mass and charge of the particle, $V_{out}$ and $\Omega = 2 \pi f $ corresponds to the amplitude and angular frequency of the AC field applied to the outer electrode. All the parameters of the trapped object and the signal applied to the PPT are contained in a single constant $A$, except for $V_{in}$ which determines the trapping position along $z$. Therefore, we can define a normalised pseudo-potential $\bar \Psi = \Psi/A$. The spatial dependence of the pseudopotential is given by $\kappa ( z,\rho,\epsilon)$, which contains Bessel functions, and simplifies to $\kappa(z,\epsilon)$ at $\rho=0$, where the trapping occurs:
\begin{equation}
\kappa (z,\epsilon) = \frac{1 - \epsilon}{\sqrt{1+ \left(\frac{a}{z}\right)^2}} - \frac{1}{\sqrt{1+ \left(\frac{b}{z}\right)^2}}+\epsilon
\label{eq:kappa}
\end{equation}

An important parameter is the dimensionless Mathieu parameter $q = \frac{8 A }{Q V_{out}}f(a,b)$ which determines the stability of the PT. Here $f(a,b)$ is the geometrical factor in units of [length]$^{-2}$ that just depends on  $a$ and $b$ (radii of the  inner and outer electrodes). In vacuum, stable solutions to the equation of motion of the trapped object exist if the condition $\mid q\mid \leq 0.9$ is satisfied \cite{Takahashi2013}. In the presence of damping, like in our ambient pressure experiments, stable trapping typically occurs at higher $q$ values \cite{izmailov1995microparticle,whitten2004high}. The pseudopotential's critical points are the minimum $z_0$ and the turning-point $z_{max}$:

\begin{eqnarray}
  \label{eq:z0}
 z_0 &=& \sqrt{\frac{b^{2} a^{4/3}(1-\epsilon)^{2/3}-a^2b^{4/3}}{b^{4/3}-a^{4/3}(1-\epsilon)^{2/3}}},\\
 z_{max} &=& \sqrt{\frac{b^2a^{4/5}(1-\epsilon)^{2/5} - a^2b^{4/5}}{b^{4/5} - a^{4/5}(1-\epsilon)^{2/5}}},
 \label{eq:zmax}
 \end{eqnarray}

which only depend on the geometry of the PCB, and in particular are independent of the damping. The trap depth is defined as $D= \Psi(z_{max})- \Psi(z_0)$.

\begin{figure}[h!]
    \centering
    \includegraphics[scale=0.37]{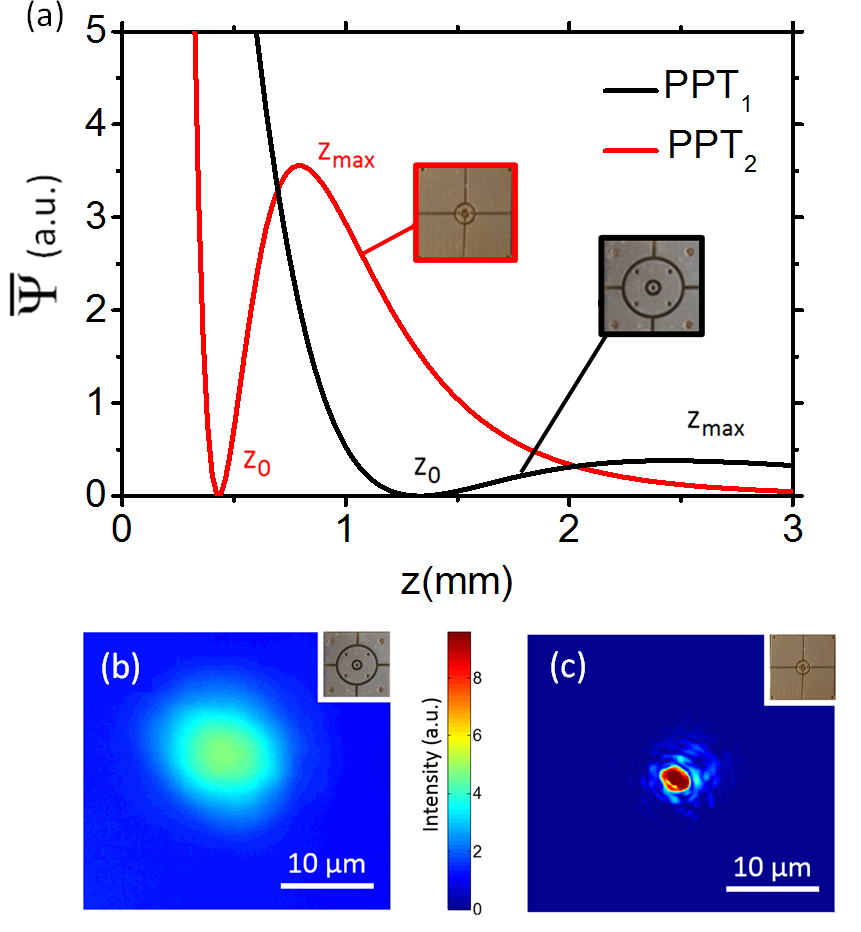}
    \caption{(a) Theoretical calculation of  the pseudopotential $\bar\Psi (z)$ for both PCBs. We found $z_{0,1}= 1.33$ mm, $z_{max,1}= 2.45$ mm, $z_{0,2}= 0.44$ mm, and $z_{max,2}= 0.78$ mm. (b-c) Temporal averaging of 372 frames for PPT$_1$ and PPT$_2$ in (b) and (c), respectively. The scale bar represents 10 $\mu$m and the color bar is the same for both images. The resolution of the camera is 0.1 $\mu$m/px.  }
    \label{confinement}
\end{figure}

To attain a higher trap confinement, we numerically determined the optimum geometrical parameters that yield to a maximum pseudo-potential depth $D$ while accounting for the manufacturing limits (imposing $a > 300 \ \mu$m). From equations (\ref{eq:pseudopot}) and (\ref{eq:kappa}), we simulated the pseudo-potential for different values of $a$ and $b$, with $a < b$. Under these constraints, the most confined pseudo-potential is obtained for $a = 0.36$ mm and $b =1.17$ mm (labeled as PPT$_2$ in the following).The simulated pseudo-potentials and their dependence on $z$ for both geometries are presented in Fig. \ref{confinement} (a). Using equations (\ref{eq:z0}) and (\ref{eq:zmax}) we determined the depth of the trap $\bar{D}= \bar{\Psi}(z_{max})- \bar{\Psi}(z_0)$. From Fig.\ref{confinement} (a),  PPT$_2$ has a trap depth of about 10 times greater than PPT$_1$.

Based on these calculations, we tested experimentally the optimised design (PPT$_2$) and compared it to the original one (PPT$_1$). We recorded through the optical objective videos of the trapped particle (acquisition area 30 $\mu$m $\times$ 30 $\mu$m at a rate of 15 frames/s) that were converted into an image sequence for analysis. Although the integration time of the camera is longer than the oscillation period of the nanoparticle and therefore it is not possible to exactly determine its spacial confinement within the trap, a time average of the image over 400 frames confirms that PPT$_2$ features much better confinement (Fig.  \ref{confinement} (b) and (c)), in agreement with the predictions shown in Fig.  \ref{confinement} (a). The image acquired from PPT$_1$ demonstrates that in this case the particle explores a large region away from the focal spot, with scattering from the particle covering the full field of view (note that the background color is different from zero). On the contrary, the image obtained with PPT$_2$ shows the nanoparticle strongly confined to a region of few microns, even suggesting the presence of an interference pattern. 

\begin{figure}[htb]
    \includegraphics[scale=0.32]{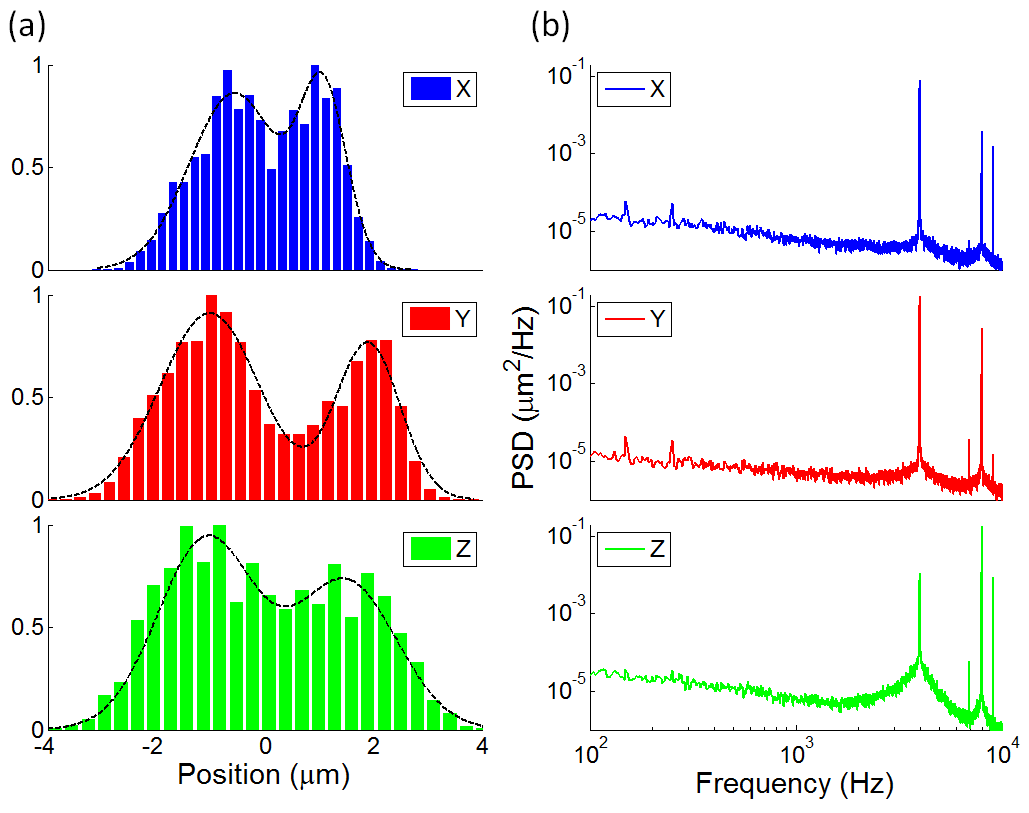}
\caption{ (a) Normalized position histogram of nanoparticle in PPT$_2$. The two maxima indicate a driven harmonic motion.  Black lines are a guide to the eye. (b) Power spectral density. We observe the driving frequency at 4 kHz and higher harmonics. }
\label{hist}
\end{figure}

To get further insight in the dynamics of the particle in PPT$_2$, we implemented a position detection scheme similar to those routinely used in optical tweezers. The trapped nanoparticle is now shined with a  $10\times$ objective (NA= 0.25) leading to a spot size $\simeq$ 2.6 $\mu$m  to ensure that we observe the full trajectory of the trapped nanoparticle. The backscattering signal from the nanoparticle was measured through the same objective and sent to a quadrant photodetector (QPD). From the time trace for each axis, we extracted the particle position histograms shown in Fig. \ref{hist} (a). The two outer maxima indicate a driven oscillation of the particle. The corresponding power spectral density (PSD) shows a dominating peak at 4 kHz--driving frequency of the PPT--as well as higher harmonics (see Fig. \ref{hist} (b)). Higher harmonics indicate a quasi-harmonic behaviour of the particle inside the trap. We estimate the confinement by fitting the low frequency part (up to 1 kHz) of the PSD to an overdamped Lorentzian, similar to what is done in optical tweezers \cite{simon2006}. Assuming the validity of the equipartition theorem of the nanoparticle in the pseudopotential, we obtain the calibration factor from volts to nanometers for each axis. Our estimated confinement is consistent with the observations in Fig. \ref{hist}(a), where the particle's trajectory is about 4 $\mu$m in all directions. 

In summary, PPTs are a non-invasive and compact technology to trap and manipulate a single nano-object in three dimensions. We have optimised  and characterised the confinement of a PPT in air which can be easily implemented on any experimental setup at low cost. The presented level of  manipulation, rotation and stability opens new possibilities for the study of nano-objects in air or vacuum, and in combination with optical traps give rise to potential applications that benefit the field of optomechanics. 

\section*{Acknowledgements}
The authors acknowledge financial support from the Fundaci\'o Privada Cellex Barcelona, the Severo Ochoa program, and CoG ERC QnanoMECA (No. 64790). I.A. acknowledges the Spanish Ministry of Education, Culture and Sport (Grant FPU014/02111).

\section*{References}
\bibliography{paultrapbio}

\end{document}